\documentclass[12pt,twoside]{article}
\usepackage{amssymb}
\usepackage{amsmath}
\usepackage{latexsym}
\usepackage{longtable}
\usepackage{epsfig}
\usepackage{graphicx,bbm,psfrag}
\graphicspath{{images/}}

\newtheorem{theorem}{Theorem}[section]
\newenvironment{proof}{{\bf Proof:}}{\hspace*{\fill}$\Box$}

\begin{document}
\begin{titlepage}
\begin{center}
\vspace*{2cm}
{\Large {\bf Superdiffusivity of the  1D Lattice Kardar-Parisi-Zhang Equation\bigskip\bigskip\\}} {\large{Tomohiro Sasamoto$^\star$ and
Herbert Spohn$^\dag$}}\bigskip\bigskip\\
 Zentrum Mathematik, TU M\"unchen,
 D-85747 Garching, Germany\\
 {$^\star$ e-mail:~sasamoto@ma.tum.de}\\
{$^\dag$ e-mail:~spohn@ma.tum.de}
\end{center}
\vspace{5cm}
\textbf{Abstract.} The continuum Kardar-Parisi-Zhang equation in one dimension is lattice
discretized in such a way that the drift part is divergence free. This allows to
determine explicitly the stationary measures. We map the lattice KPZ
equation to a bosonic field theory which has a cubic anti-hermitian nonlinearity.
Thereby it is established that the stationary two-point function spreads superdiffusively.
\end{titlepage}

\section{Introduction}

Oversimplified models of surface growth have remained of interest,
in particular their one-dimensional version. In very general terms,
one investigates a conserved field with a stochastic dynamics which
does not satisfy the condition of detailed balance. Mostly models
with discrete space and discrete heights have been in focus, as e.g.
the asymmetric exclusion processes (ASEP), also its totally
asymmetric version (TASEP), and the polynuclear growth (PNG) model.
We refer to \cite{Joh06,Spo06,Sas07} for reviews. Some features of
these models are related to integrable systems and random matrix
theory, which has become an independent motivation for intense
study. There are unexpected connections to random tilings, as the
Aztec diamond, and to crystal shapes, somewhat less surprising to
directed last passage percolation, also known as directed polymer in
a random medium. For several features of ASEP, TASEP,  and PNG exact
solutions can be obtained through a rather intricate asymptotic
analysis. In particular, the stationary two-point function has been
computed in the scaling limit \cite{PS04,FS06}. Except for model
dependent scale factors one finds the same scaling function in the
various models, supporting, at least partially, the universality
hypothesis for growth models with local deposition rules and
neglected surface diffusion.

In a seminal paper, Kardar, Parisi, and Zhang \cite{KPZ86} proposed
a continuum equation for surface growth. In one dimension their
equation reads
\begin{equation}\label{1.1}
\partial_t h(x,t)= \tfrac{1}{2} \lambda_\mathrm{b}(\partial_x
h(x,t))^2 + \nu_\mathrm{b}\partial^2_x h(x,t) +
\sqrt{D_\mathrm{b}}\xi (x,t)\,.
\end{equation}
Here $h$ is the height function over $\mathbb{R}$ at time $t$,
$\lambda_\mathrm{b}$ is the strength of the nonlinear growth
velocity, $\nu_\mathrm{b}$ is the strength of the local smoothening,
and $\xi$ is normalized Gaussian space-time white noise,
$\langle\xi(x,t)\xi(x',t')\rangle=\delta(x-x') \delta(t-t')$. The
slope $u=\partial_x h$ then satisfies the stochastic Burgers
equation,
\begin{equation}\label{1.2}
\partial_t u= \tfrac{1}{2} \lambda_\mathrm{b}\partial_x
u^2 + \nu_\mathrm{b}\partial^2_x u +
\sqrt{D_\mathrm{b}}\partial_x\xi\,.
\end{equation}
Note that this relation is a special property of one dimension.

It is expected that Eq. (\ref{1.2}) is in the same universality
class as ASEP and PNG. The most convincing evidence comes from the
numerical simulation of the discretized version of (\ref{1.1}) with
1024 lattice sites \cite{KS04}. The stationary two-point function is
computed and good agreement  is obtained with the exact solution
from TASEP and PNG over a substantial range of wave vectors.

On the theoretical side, one notes that the noise in (\ref{1.2}) is
very singular. To cope with this difficulty, one worked out approach
\cite{Be97} is to make Eq. (\ref{1.1}) linear through the Cole-Hopf
transformation
\begin{equation}\label{1.2a}
   Z(x,t)= \exp [(\lambda_\mathrm{b}/ 2\nu_\mathrm{b})h(x,t)]\,.
\end{equation}
Then the ``partition function'' $Z(x,t)$ satisfies
\begin{equation}\label{1.2b}
\partial_t Z(x,t)= \nu_\mathrm{b}\partial^2_x
Z(x,t) + \frac{\lambda_\mathrm{b}\sqrt{D_\mathrm{b}}}{ 2\nu_\mathrm{b}}\xi(x,t) Z(x,t)\,,
\end{equation}
with initial conditions $Z(x,0)>0$. Being linear, one can give sense
to $Z(x,t)$ as a stochastic process with continuous sample paths. In particular, $Z(x,t)>0$ with
probability 1 and one \textit{defines} $h(x,t)$ through Eq.
(\ref{1.2a}). In fact, this type of solution of the KPZ equation can
be recovered by a suitable continuum limit of the ASEP with a
properly chosen weak asymmetry.

An alternative approach is to regularize the noisy Burgers equation
(\ref{1.2}). Depending on the point view there are then two choices.
The fluid dynamics camp regards (1.2) as oversimplified large scale
randomly stirred Navier-Stokes equations and studies small scale
properties of the solution, see \cite{Got98,E00} out of a large body
of literature. But then it is natural to keep the $\delta$-function
in time and to replace the $\delta$-function in space by a
smoothened version. On the other hand the surface growth community
knows that the lattice structure of the solid defines a smallest
length scale. Thus one discretizes the KPZ equation, as done without
further ado in any numerical simulation, and studies its properties
on large scales. In relation to ASEP and PNG it is natural to follow
the second route and to discretize $\mathbb{R}$ as $\delta\mathbb{Z}$
with lattice constant $\delta>0$. The field variables are denoted by
$u_j(t)\in\mathbb{R}$, $j\in\mathbb{Z}$, as the lattice version of
$u(x,t)$ in (\ref{1.2}). Then the lattice KPZ equation, studied in
this paper, reads
\begin{eqnarray}\label{1.3}
&&\hspace{-38pt}\frac{d}{dt}u_j=
\tfrac{1}{2}(\lambda_\mathrm{b}/3\delta)(u^2_{j+1} + u_ju_{j+1}
- u_{j-1}u_j-u^2_{j-1})\nonumber\\
&&\hspace{2pt}+(\nu_\mathrm{b}/\delta^2) (u_{j+1}-2u_j+u_{j-1})+
(D_\mathrm{b}/\delta)^{1/2}(\xi_j-\xi_{j-1})\,,\;j\in\mathbb{Z}\,.
\end{eqnarray}
Here $\{\xi_j,j\in\mathbb{Z}\}$ is a collection of independent and normalized
white noises. Below we will determine a family of stationary measures of
(\ref{1.3}). We consider then the space-time stationary solutions to
(\ref{1.3}) and the object of prime interest will be stationary
two-point function
\begin{equation}\label{1.4}
S_\rho(j,t)=\langle u_j(t) u_0(0)\rangle_\rho-\langle u_0(0)\rangle_\rho^2
\end{equation}
at average slope $\rho = \langle u_j(t)\rangle_\rho$.

Each one of the three terms on the right hand side of  (\ref{1.3}) converges
as $\delta\to 0$ to the corresponding term in
(\ref{1.2}). To obtain the scale invariant theory, however, one has
to adjust the ``coupling constants'' in (\ref{1.3}) in such a way
that $S(j,t)$ converges to a nontrivial limit. This is the reason for the subscript
b, which stands for bare coupling parameters.

To provide a brief summary, in Section \ref{sec2} we rewrite
(\ref{1.3}) as a bosonic field theory with a cubic anti-hermitian
nonlinearity, where for simplicity we consider only the case
of zero slope, $\rho=0$. As explained in Section  \ref{sec3}, restricting the bosonic occupation
variables to 0,1, i.e. restricting to hard core bosons, yields
precisely the stationary ASEP at density 1/2. The bosonic
representation is used in the study of the spreading of
$S_{\rho=0}(j,t)$. $S_{\rho=0}(j,t)$ is even in $j$ and the
normalized variance reads
\begin{equation}\label{1.5}
\textrm{Var}(t)=\chi^{-1} \sum_{j\in\mathbb{Z}} j^2
S_{\rho=0}(j,t)\,,\quad \chi=\sum_{j\in\mathbb{Z}}
S_{\rho=0}(j,0)\,.
\end{equation}
In Section \ref{sec5} we will establish the upper and lower bounds
as
\begin{equation}\label{1.6}
t^{5/4} \leq \textrm{Var}(t)\leq t^{3/2}
\end{equation}
as $t\to\infty$. We then develop an iterative scheme based on the
relaxation time approximation and recover the KPZ prediction of
$\textrm{Var}(t) \cong t^{4/3}$. Very recently J. Quastel announced
a proof of the following bounds,
\begin{equation}\label{1.7}
c_- t^{4/3} \leq \mathrm{Var}_{\mathrm{CH}}(t) \leq c_+ t^{4/3}
\end{equation}
for large $t$ and suitable constants $0< c_- < c_+$ \cite{Qu09}. Here
$\mathrm{Var}_{\mathrm{CH}}(t)$ is computed from the variance of
$\log Z(x,t)$ with $Z(x,t)$ the Cole-Hopf solution of (\ref{1.2b})
with initial data $Z(x,0)$ such that $\log Z(x,0)$ is two-sided Brownian motion in $x$.
In the final section the continuum approximation is discussed and
contrasted with the well studied case of (\ref{1.2}) for
$D_\mathrm{b}=0$, $\nu_\mathrm{b}\to 0$, and white noise initial
data.

\section{The lattice KPZ equation}\label{sec2}
\setcounter{equation}{0}

To simplify notation we set $\lambda_0=\lambda_\mathrm{b}/\delta$,
$\nu_0=\nu_\mathrm{b}/\delta^2$, $D_0=D_\mathrm{b}/\delta$, and $\alpha = \nu_0/D_0$.

The lattice KPZ equation (\ref{1.3}) conserves the slope field $u$,
which becomes manifest through introducing the current function
\begin{eqnarray}\label{2.1}
&&w_j=\tfrac{1}{6} \lambda_0 (u^2_{j} +u_j u_{j+1} + u^2_{j+1}) +
\nu_0(u_{j+1}-u_j)\, \nonumber\\
&&\hspace{16pt}=\tilde{w}_j + \nu_0(u_{j+1}-u_j)\,.
\end{eqnarray}
Then
\begin{equation}\label{2.2}
\frac{d}{dt}u_j= w_j-w_{j-1} + \sqrt{D_0}(\xi_j-\xi_{j-1})\,,\quad
j\in \mathbb{Z}\,.
\end{equation}
In principle, there are many possibilities to discretize $\partial_x
u(x)^2$ in (\ref{1.2}). Our choice is singled out by the facts  (i)
the discretization involves only nearest neighbors and (ii) the
drift term is a divergence free vector field, i.e.
\begin{equation}\label{2.2a}
\sum_{j \in \mathbb{Z}}\partial_j(w_j-w_{j-1})=0\,,
\end{equation}
where $\partial_j=\partial/\partial u_j$. For this particular
discretization one can compute explicitly the invariant measures.
(\ref{2.1}) and (\ref{2.2}) was first proposed in \cite{KS}, see
also \cite{LS98} for a more recent study.

For a while we study (\ref{2.2}) on the ring $[1,\ldots,N]$, i.e.
\begin{equation}\label{2.3}
u_{N+j}=u_j\quad \mathrm{and} \quad \xi_{N+j}=\xi_j\,.
\end{equation}
The generator for the diffusion process (\ref{2.2}) is then
\begin{equation}\label{2.4}
L_N= \sum^N_{j=1}\big[(w_j-w_{j-1}) \partial_j+\tfrac{1}{2}
D_0(\partial_j-\partial_{j-1})^2\big]\,.
\end{equation}
 Noting the identity
\begin{equation}\label{2.5}
\sum^N_{j=1}(w_j-w_{j-1})u_j=0\,,
\end{equation}
one verifies that, for every $\rho\in\mathbb{R}$, the independent Gaussians
\begin{equation}\label{2.6}
\prod^N_{j=1}\big\{(\alpha/\pi)^{1/2}\exp[-\alpha(u_j-\rho)^2]\big\}=
\big(\psi^N_{0,\rho}(\underline{u})\big)^2
\end{equation}
are invariant for (\ref{2.4}), i.e. for every smooth function $f$ on
configuration space it holds
\begin{equation}\label{2.6a}
\int_{\mathbb{R}^N} \mathrm{d}\underline{u} \psi^N_{0,\rho} (\underline{u})^2
L_N f(\underline{u})=0\,.
\end{equation}
Note that the stationary measure does not depend on $\lambda_0$. Let
us denote the average with respect to $(\psi^N_{0,\rho})^2$ by
$\langle\cdot\rangle_{\rho,N}$, $\langle 1\rangle_{\rho,N}=1$. Then
\begin{equation}\label{2.7}
\langle u_j\rangle_{\rho,N}=\rho
\end{equation}
is the average slope,
\begin{equation}\label{2.6b}
\langle u^2_j\rangle_{\rho,N}- \langle u_j\rangle^2_{\rho,N}=\frac{1}{2\alpha}
\end{equation}
the variance, and
\begin{equation}\label{2.8}
\langle w_j\rangle_{\rho,N}=\tfrac{1}{6}\lambda_0
\big(\alpha^{-1}+3\rho^2\big)=j(\rho)
\end{equation}
the average current. For simplicity we will consider only the slope 0
case, $\rho=0$, and hence omit the index $\rho$.

We switch from the generator $L_N$ to the
hamiltonian $H_N$ through the ground state transformation
\begin{equation}\label{2.9}
\psi^N_0 L_N(\psi^N_0)^{-1}=-H_N\,.
\end{equation}
Then
\begin{equation}\label{2.10}
H_N=\tfrac{1}{2}D_0\sum^N_{j=1}\big(-(\partial_j-\partial_{j-1})^2+\alpha^2(u_j-u_{j-1})^2- 2\alpha\big)
-\sum^N_{j=1}(\tilde{w}_j-\tilde{w}_{j-1})\partial_j\,.
\end{equation}
We introduce the standard annihilation/creation operators at site
$j$ through
\begin{equation}\label{2.11}
a_j=\frac{1}{\sqrt{2\alpha}}(\alpha u_j+\partial_j)\,,\quad
a^\ast_j=\frac{1}{\sqrt{2\alpha}}(\alpha u_j-\partial_j)\,.
\end{equation}
Note that they satisfy the canonical commutation relations
\begin{equation}\label{2.12}
[a_i,a^\ast_j]=\delta_{ij}\,.
\end{equation}
Then
\begin{equation}\label{2.13}
H_N=\tilde{H}_{0,N}+\lambda_0(\tilde{A}_N^\ast-\tilde{A}_N)\,,
\end{equation}
where
\begin{equation}\label{2.14}
\tilde{H}_{0,N}= \nu_0\sum^N_{j=1} (a_{j+1}-a_j)^\ast
(a_{j+1}-a_j)
\end{equation}
and
\begin{equation}\label{2.15}
\tilde{A}_N= (3\cdot2^{3/2})^{-1}\alpha^{-1/2} \sum^N_{j=1} \big(a_j
a^\ast_{j+1} a_{j+1} + a_j a_j a^\ast_{j+1} -a^\ast_{j} a_{j+1}
a_{j+1}-a^\ast_{j}  a_{j}a_{j+1}\big)\,.
\end{equation}

We want to compute the propagator $\exp[-t H_N]$, $t\geq 0$. To reduce the number
of parameters we rescale time as
$t\rightsquigarrow \nu_0t$. Then the prefactor of
$H_{0,N}$ becomes one. We also introduce the coupling constant
\begin{equation}\label{2.15a}
\lambda=(3\cdot2^{3/2})^{-1}\lambda_0\alpha^{-1/2}\nu_0^{-1}\,.
\end{equation}
Then
\begin{equation}\label{2.15b}
H_N=H_{0,N}+\lambda(A^\ast_N-A_N)\,,
\end{equation}
where
\begin{equation}\label{2.15c}
H_{0,N}=\sum^N_{j=1} (a_{j+1}-a_j)^\ast(a_{j+1}-a_j)\,,
\end{equation}
\begin{equation}\label{2.15d}
A_N=\sum^N_{j=1} \big(a_j a^\ast_{j+1} a_{j+1} + a_j a_j
a^\ast_{j+1} -a^\ast_{j} a_{j+1} a_{j+1}-a^\ast_{j}
a_{j}a_{j+1}\big) \,.
\end{equation}

It is convenient to  use the \textit{Fock space} representation of
the bosonic field $a_j, a^\ast_{j}$. Physically this corresponds to
a study of local excitations away from the Gaussian stationary
measure, as e.g. encoded by the two-point function $S(j,t)$. For
$N=\infty$ the Fock space $\mathfrak{F}$ is over
$\ell_2=\ell_2(\mathbb{Z})$ with the $n$-particle space
$\mathfrak{F}_n=(\ell_2)^{\otimes n}_{\mathrm{sym}}$ and
\begin{equation}\label{2.20}
   \mathfrak{F}=\bigoplus^\infty_{n=0} \mathfrak{F}_n\,.
\end{equation}
An element $f\in\mathfrak{F}$ is a sequence
$\{f_0,f_1,\ldots,f_n,\ldots\}$. $f_n:\mathbb{Z}^n\to\mathbb{C}$ and
$f_n$ is symmetric in its arguments, $f_0\in\mathbb{C}$. The scalar
product on $\mathfrak{F}_n$ is
\begin{equation}\label{2.21}
\langle f,g\rangle_n =\sum_{(x_1,\ldots,x_n)\in\mathbb{Z}^n}
f_n(x_1,\ldots,x_n)^\ast g_n(x_1,\ldots,x_n)\,.
\end{equation}
$f\in \mathfrak{F}$ if and only if
\begin{equation}\label{2.22}
\langle f,f\rangle =|f_0|^2 +\sum^\infty_{n=1} \langle
f_n,f_n\rangle_n < \infty\,.
\end{equation}
$\mathfrak{F}_n$ is the $n$-particle subspace. The Fock vacuum is
the vector $\Omega=(1,0,0,\ldots)$.

The $\{a_j,a^\ast_j, j\in\mathbb{Z}\}$ are the usual bosonic
annihilation/creation operators on $\mathfrak{F}$, which are defined
through
\begin{equation}\label{2.22a}
(a_j f_{n+1}) (x_1,\ldots,x_n)= \sqrt{n+1} f_{n+1}
(j,x_1,\ldots,x_n)
\end{equation}
and
\begin{equation}\label{2.22b}
(a^\ast_j f_{n-1}) (x_1,\ldots,x_n)= \frac{1}{\sqrt{n}}
\sum^n_{\ell=1} \delta(x_\ell -j) f_{n-1}
(x_1,\ldots,\hat{x}_\ell,\ldots,x_n)
\end{equation}
with the convention that $\hat{x}_\ell$ means the omission of
$x_\ell$ from the configuration $(x_1,$ $\ldots,x_n)$, see e.g. \cite{S66}. From
(\ref{2.15b}) -- (\ref{2.15d}) in the limit $N \to \infty$, we
arrive at
\begin{equation}\label{2.23}
H=H_0 + \lambda(A^\ast-A)\,.
\end{equation}
Explicitly
\begin{equation}\label{2.24}
H_0=\sum_{j\in\mathbb{Z}} (a_{j+1} -a_j)^\ast (a_{j+1} -a_j)
\end{equation}
and
\begin{equation}\label{2.24a}
A=\sum_{j\in\mathbb{Z}}\big(a_j a^\ast_{j+1} a_{j+1} + a_j a_j
a^\ast_{j+1} -a^\ast_{j} a_{j+1} a_{j+1}-a^\ast_{j}
a_{j}a_{j+1}\big)
\end{equation}
as operators on Fock space.

$H_0\upharpoonright\mathfrak{F}_n$ is the Laplacian and corresponds
to $n$ independent random walks on $\mathbb{Z}$ with nearest
neighbor hopping of rate 1. Clearly, $H_0=H^\ast_0$, while
$A^\ast-A$ is antisymmetric, $(A^\ast-A)^\ast=-(A^\ast-A)$. Using
(\ref{2.22a}), (\ref{2.22b}), the action of $A$ on Fock space is
represented by
\begin{eqnarray}\label{2.22c}
 &&\hspace{-25pt} A f_{n+1}(x_1,\ldots,x_n)=\sqrt{n+1}
\sum^n_{\ell=1}\big(f_{n+1}(x_1,\ldots,x_n,x_\ell -1)\nonumber\\
&&\hspace{2pt}-f_{n+1}(x_1,\ldots,x_n,x_\ell +1)+
f_{n+1}(x_1,\ldots,\hat{x}_\ell,\ldots,x_n,x_\ell -1,x_\ell
-1)\nonumber\\
&&\hspace{12pt}-f_{n+1}(x_1,\ldots,\hat{x}_\ell,\ldots,x_n,x_\ell
+1,x_\ell +1)\big)\,.
\end{eqnarray}

To have an
example, let us consider the two-point function at slope zero,
\begin{equation}\label{2.25a}
S(j,t)=\mathbb{E}(u_j(t) u_0(0))\,.
\end{equation}
Here $N=\infty$ and the expection $\mathbb{E}(\cdot)$ refers to the
solution of (\ref{2.1}), (\ref{2.2}) with $u_j(0)$ distributed as
independent Gaussians with mean 0 and variance $1/2\alpha$. In Fock
space $S(j,t)$ is expressed through
\begin{equation}\label{2.25}
S(j,t)= \frac{1}{2\alpha}\langle \Omega, a_0 \mathrm{e}^{-\nu_0tH} a^\ast_j \Omega\rangle\,,\quad
t\geq 0\,.
\end{equation}
Note that  by the reflection $u_j \leadsto -u_{-j}$ and by time stationarity, one concludes
\begin{equation}\label{2.26}
S(j,t)= S(-j,t)\,,\quad S(j,t)= S(j,-t)\,,\quad t\geq 0\,.
\end{equation}

Computationally it is often more convenient to switch to Fourier
space. For $f:\mathbb{Z}\to\mathbb{C}$ we define the Fourier
transform
\begin{equation}\label{2.27}
\widehat{f}(k)=\sum_{j\in\mathbb{Z}} \mathrm{e}^{-i 2\pi kj} f(j)
\end{equation}
with inverse
\begin{equation}\label{2.28}
f(j)= \int^1_0 \mathrm{d}k  \mathrm{e}^{i 2\pi kj} \widehat{f}(k)\,.
\end{equation}
Thus the one-particle space is now $L^2([0,1],\mathrm{d}k)$. We set
$\mathbb{T}=[0,1]$ as first Brillouin zone. Correspondingly
\begin{equation}\label{2.29}
a(k)=\sum_{j\in\mathbb{Z}} \mathrm{e}^{-i 2\pi kj}a_j\,.
\end{equation}
Note that
\begin{equation}\label{2.30}
[a(k), a(k')^\ast]=\delta(k-k')\,,
\end{equation}
which means that the corresponding map between Fock spaces is
unitary. In Fourier representation it holds
\begin{equation}\label{2.31}
H_0= \int_\mathbb{T} \mathrm{d}k \omega(k) a(k)^\ast a(k)\,,\quad
\omega(k)=2(1-\cos(2\pi k))
\end{equation}
and
\begin{eqnarray}\label{2.32}
 &&\hspace{-35pt}  A=- i\int_{\mathbb{T}^3}\mathrm{d}k_1 \mathrm{d}k_2
\mathrm{d}k_3 \delta (k_1-k_2-k_3) \big(2
\sin(2\pi k_1)\nonumber\\
&&\hspace{32pt}+ \sin(2\pi k_2)+ \sin(2\pi k_3)\big) a(k_1)^\ast a(k_2) a(k_3)\,.
\end{eqnarray}

\section{Relation to the ASEP at density 1/2}\label{sec3}
\setcounter{equation}{0}

The partially asymmetric simple exclusion process (ASEP) is a
stochastic particle system on $\mathbb{Z}$ with hard exclusion, i.e.
the occupation variable $\eta_j$ at site $j$ takes only the values
0,1. The state space is $\{0,1\}^\mathbb{Z}$. Particles hop to the
right neighbor with rate $1+p$ and to the left neighbor with rate $1-p$,
$|p|\leq 1$, provided the destination site is empty. Therefore the
generator, $L_{\mathrm{AS}}$, reads
\begin{equation}\label{3.1}
L_{\mathrm{AS}} f(\eta)=
\sum_{j\in\mathbb{Z}}\big((1+p)\eta_j(1-\eta_{j+1})
+(1-p)(1-\eta_j)\eta_{j+1}\big) \big(f(\eta^{jj+1})-f(\eta)\big)\,,
\end{equation}
where $\eta^{jj+1}$ denotes the configuration $\eta$ with the
occupations at sites $j$ and ${j+1}$ interchanged. The Bernoulli
measures are invariant under $L_{\mathrm{AS}}$. We consider only the
stationary process with Bernoulli 1/2.

We expand $L_{\mathrm{AS}}$ in the natural basis for the Bernoulli
measure with density 1/2 \cite{LQSY}. The basis functions are
labelled by finite subsets, $\Lambda$, of $\mathbb{Z}$ and are of
the form
\begin{equation}\label{3.1a}
\psi_{\Lambda}(\eta) = \prod_{j \in \Lambda} (2\eta_j - 1)\,.
\end{equation}
In particular, with $\langle\cdot\rangle_{1/2}$ denoting average
over Bernoulli 1/2,
\begin{equation}\label{3.2}
 \langle\psi_{\Lambda_1}\psi_{\Lambda_2}\rangle_{1/2} = \delta(\Lambda_1,\Lambda_2)\,.
\end{equation}
A general function, $v$, is represented by
\begin{equation}\label{3.2a}
v(\eta) =  \sum_{\Lambda\subset\mathbb{Z}}
\hat{v}(\Lambda)\psi_\Lambda(\eta)\,,\quad
\sum_{\Lambda\subset\mathbb{Z}} | \hat{v}(\Lambda)|^2 <\infty\,,
\end{equation}
the sum being over all finite subsets of $\mathbb{Z}$. In this
basis $L_\mathrm{AS}$ is represented by
\begin{equation}\label{3.3}
L_\mathrm{AS}=-H_\mathrm{AS}\,,\quad H_\mathrm{AS}=
\mathcal{S}+p(\mathcal{A}^\ast-\mathcal{A})\,.
\end{equation}
Here $\mathcal{S}=\mathcal{S}^\ast$ and
\begin{equation}\label{3.3a}
\mathcal{S} \hat{v}(\Lambda)=
-\sum_{x\in\mathbb{Z}}\big(\hat{v}(\Lambda_{x,x+1})-\hat{v}(\Lambda)\big)\,,
\end{equation}
where $\Lambda_{x,x+1}$ is obtained from $\Lambda$ by exchanging the
occupancies at $x$ and $x+1$. To define $\mathcal{A}$ and
$\mathcal{A}^\ast$ we introduce the outer left boundary of
$\Lambda$, $\ell(\Lambda)= \{x|x\notin \Lambda, x+1\in\Lambda\}$,
and the outer right boundary of
$\Lambda,r(\Lambda)=\{x|x\notin\Lambda, x-1\in\Lambda\}$.
Correspondingly, the inner left and inner right boundary of
$\Lambda$ are $\bar{\ell}(\Lambda)= \{x|x\in\Lambda,
x-1\notin\Lambda\}$, $\bar{r}(\Lambda)=\{x|x\in\Lambda,
x+1\notin\Lambda\}$. Then
\begin{eqnarray}\label{3.3b}
 &&\hspace{-15pt}\mathcal{A}\hat{v}(\Lambda)=\sum_{x\in\ell(\Lambda)}
\hat{v}(\Lambda\cup\{x\})-\sum_{x\in r(\Lambda)} \hat{v}(\Lambda\cup\{x\})\,,\nonumber\\
&&\hspace{-15pt}\mathcal{A^\ast}\hat{v}(\Lambda)=\sum_{x\in\bar{\ell}(\Lambda)}
\hat{v}(\Lambda\setminus\{x\})-\sum_{x\in \bar{r}(\Lambda)}
\hat{v}(\Lambda\setminus\{x\})\,.
\end{eqnarray}

$L_{\textrm{AS}}$ is the generator in the number space
representation. To be able to compare with $H$ one still has to
transform unitarily to Fock space representation. For this purpose
let $|\Lambda|=n$, $\Lambda=\{x_1,\ldots,x_n\}$. Out of $\hat{v}(\Lambda)$
we construct $f_n\in\mathfrak{F}_n$ by
\begin{equation}\label{3.3c}
f_n(x_1,\ldots,x_n)=
\begin{cases}
(1/\sqrt{n!})\hat{v}(\{x_1,\ldots,x_n\}) & (x_1,\ldots,x_n)\textrm{ has no coinciding points}\,, \\
 0 & \textrm{otherwise}\,.
\end{cases}
\end{equation}
We observe that
\begin{eqnarray}\label{3.3d}
&&\hspace{-46pt}\langle f_n,f_n \rangle_n =
\sum_{\underline{x}\in\mathbb{Z}^n} |f_n(x_1,\ldots,x_n)|^2
\nonumber \\ &&= \frac{1}{n!}
\sum_{\underline{x}\in\mathbb{Z}^n,x_\ell \neq x_m,\;\mathrm{all}\;
\ell\neq m} | \hat{v}(\{x_1,\ldots,x_n\})|^2 =
 \sum_{\Lambda\subset\mathbb{Z},|\Lambda|=n} |\hat{v}(\Lambda)|^2\,.
\end{eqnarray}
Hence the map defined by (\ref{3.3c}) is unitary. The set of all
$\hat{v}$'s such that the right hand side of
 (\ref{3.3d}) is finite
defines the subspace $P\mathfrak{F}_n$ of $\mathfrak{F}_n$.
$P\mathfrak{F}_n$ consists of functions which vanish whenever their
argument $(x_1,\ldots,x_n)$ has coinciding points. $P$ is a projection operator on $\mathfrak{F}_n$
and $P$ as an operator on $\mathfrak{F}$ is defined
through its action on each subspace $\mathfrak{F}_n$.

Now, if $f_{n+1}\in P\mathfrak{F}_{n+1}$, then in the expression (\ref{2.22c})
for $A$ the
two last summands vanish. $A$ carries the prefactor $\sqrt{n+1}$ which
is balanced by the
normalization $1/\sqrt{n!}$ in (\ref{3.3c}). Thus we conclude that, for $ \lambda=p$,

\begin{equation}\label{3.3e}
PHP=H_\mathrm{AS}
\end{equation}
on Fock space. Restricting the lattice KPZ hamiltonian to the
subspace $P\mathfrak{F}$ results in the hamiltonian of the
asymmetric simple exclusion process. Note that $L_{\mathrm{AS}}$ is the generator of a Markov jump
process only for $|p|\leq 1$. On the other hand, $H$ is defined for all $\lambda$ and thus
(\ref{3.3e}) provides an ``analytic continuation'' of $L_{\mathrm{AS}}$.

It would be most useful to have comparison inequalities between
lattice KPZ and ASEP. For the symmetric part of the lattice KPZ
hamiltonian, the quadratic form on $\mathfrak{F}_n$ reads
\begin{equation}\label{3.3f}
\langle f_n, H_0
f_n\rangle_n=\sum^n_{\ell=1}\sum_{x_1,\ldots,x_n\in\mathbb{Z}} |
f_n(x_1,\ldots,x_\ell+1,\ldots,x_n)-f_n(x_1,\ldots,x_n)|^2\,.
\end{equation}
Thus $H_0$ restricted to $\mathfrak{F}_n$   is the Laplacian on $\mathbb{Z}^n$. On  $\mathfrak{F}_n$ the
projected operator $PH_0P$ corresponds to the Neumann restriction of the Laplacian (\ref{3.3f})
to the set $\{\underline{x}\in\mathbb{Z}^n|x_\ell\neq x_m$ for all
pairs $\ell\neq m\}$. Hence for the symmetric part it holds
\begin{equation}\label{3.3g}
H_0\geq PH_0P\,
\end{equation}
on  $\mathfrak{F}$.

On the other hand, there seems to be no such simple relation for the
asymmetric part. In \cite{LQSY} the authors approximate the ASEP by
a model without hard exclusion, in such a way that its restriction
to the range of $P$ agrees with the ASEP. Restricted to the subspace $\bigoplus^n_{m=0}
\mathfrak{F}_m$, they bound the  resolvent of
one hamiltonian in terms of the resolvent of the other hamiltonian, and vice versa.
Their bounds are not uniform in $n$. Since the
approximation in \cite{LQSY} has some similarity to the lattice KPZ hamiltonian,
we expect that their bounds would carry over to  $H$ from
(\ref{2.23}).

\section{Variance of the two-point function for large $t$}\label{sec4}
\setcounter{equation}{0}

Since, by the conservation law,
\begin{equation}\label{4.2}
 \chi=\sum_{j\in\mathbb{Z}} S(j,0) = \sum_{j\in\mathbb{Z}} S(j,t)\,,\quad  \chi = \frac{1}{2\alpha}\,,
 \end{equation}
and since, by (\ref{2.26}),
\begin{equation}\label{4.2a}
\sum_{j\in\mathbb{Z}}
j S(j,t) = 0\,,
\end{equation}
it is natural to introduce the normalized variance
\begin{equation}\label{4.1}
\mathrm{Var}(t)=\chi^{-1}\sum_{j\in\mathbb{Z}} j^2 S(j,t)\,.
\end{equation}
We first represent $S(j,t)$ through the transition probability
$\mathrm{e}^{Lt}$, where $L$ is the generator $L_N$ from (\ref{2.4})
with the sum over all $j\in\mathbb{Z}$. Then
\begin{equation}\label{4.2b}
S(j,t)= \mathbb{E}(u_0(0) u_j(t))= \langle u_0 \mathrm{e}^{L
t}u_j\rangle\,.
\end{equation}
The expectation $\mathbb{E}$ has been defined below (\ref{2.25a})
and $\langle\cdot\rangle$ refers to the average over the initial
data, i.e. over independent Gaussians with mean $0$ and variance $1/2\alpha$. $u_0$ and
$\mathrm{e}^{Lt}u_j$ are functions on configuration space. The
conservation law is used again to partially integrate twice,
\begin{eqnarray}\label{4.1a}
 &&\hspace{-38pt}\chi\mathrm{Var}(t)=\sum_{j\in\mathbb{Z}}j^2
 \langle u_0 \mathrm{e}^{Lt}u_j\rangle\nonumber\\
&&\hspace{5pt}=\sum_{j\in\mathbb{Z}}j^2
 \big( \langle u_0 u_j\rangle + \int^t_0 \mathrm{d}s \langle u_0 \mathrm{e}^{Ls}L
u_j\rangle\big)\nonumber\\
&&\hspace{5pt}=\int^t_0 \mathrm{d}s \sum_{j\in\mathbb{Z}}j^2
 \langle u_0 \mathrm{e}^{Ls} (w_j -w_{j-1})\rangle\nonumber\\
&&\hspace{5pt}=\int^t_0 \mathrm{d}s \sum_{j\in\mathbb{Z}}(-2j-1)\Big(
 \langle u_0 w_j\rangle +\int^s_0 \mathrm{d}s'\langle(L^\ast u_0)
\mathrm{e}^{Ls'}w_j
\rangle\Big)\nonumber\\
&&\hspace{5pt}=2t \nu_0
 \langle u_0^2\rangle + 2\int^t_0 \mathrm{d}s \int^s_0 \mathrm{d}s'
\sum_{j\in\mathbb{Z}}\big(\langle( w_0 - \langle
w_0\rangle)\mathrm{e}^{Ls'} (w_j- \langle w_j\rangle)\rangle
\big)\,.
\end{eqnarray}
It is convenient to switch to Laplace transform. Then
\begin{eqnarray}\label{4.1b}
 &&\hspace{-32pt}\chi\mathrm{\hat{V}ar}(\zeta)=\int^\infty_0  \mathrm{d}t \mathrm{e}^{-\zeta \nu_0 t}
\chi\mathrm{Var}(t)\nonumber\\
&&\hspace{15pt}= 2\zeta^{-2} \nu_0^{-1} \langle u^2_0\rangle + 2(\nu_0
\zeta)^{-2} \sum_{j\in\mathbb{Z}}\langle( w_0 - \langle w_0\rangle)\frac{1}{\nu_0 \zeta-L}
(w_j -\langle w_j\rangle)\rangle
\end{eqnarray}
for $\zeta>0$.

On the other hand, the KPZ scaling theory asserts that
\begin{equation}\label{4.3}
S(j,t) \cong  \chi(2\lambda_0^2 \chi t^2)^{-1/3} f_{\mathrm{KPZ}}
\big((2\lambda_0^2 \chi t^2)^{-1/3} j\big)
\end{equation}
with $\chi$ defined in (\ref{4.2}), the renormalized coupling constant
$j''(0) = \lambda_0$, and  the universal stationary
KPZ scaling function $f_{\mathrm{KPZ}}$. The validity of (\ref{4.3}) has been proved for
the stationary PNG model  \cite{PS04} and for the stationary TASEP
 \cite{FS06}. We refer to these papers for the definition of  $f_{\mathrm{KPZ}}$.
In particular,
\begin{equation}\label{4.4}
\mathrm{Var}(t) =  (2\lambda_0^2 \chi t^2)^{2/3} \langle x^2\rangle_{\mathrm{KPZ}}
\end{equation}
for $t\to\infty$ with $ \langle x^2\rangle_{\mathrm{KPZ}} = \int \mathrm{d}xx^2 f_\mathrm{KPZ}(x)= 0.510523\ldots$ a model
independent number. Hence
\begin{equation}\label{4.4a}
 \mathrm{\hat{V}ar}(\zeta)=( 2\lambda^2_0 \chi)^{2/3} \Gamma(7/3) (\nu_0 \zeta)^{-7/3}
   \langle x^2\rangle_{\mathrm{KPZ}}
\end{equation}
for $\zeta\to0$. Comparing (\ref{4.4a}) and (\ref{4.1b}) yields the
prediction for the small $\zeta$ behavior of the resolvent in
(\ref{4.1b}). But before we have to reexpress this resolvent in Fock
space.

Let us define the total momentum operator, $P_{\mathrm{tot}}$, as
\begin{equation}\label{4.5}
   P_{\mathrm{tot}}= 2\pi \int_\mathbb{T}\mathrm{d}k k a(k)^\ast a(k)\,.
\end{equation}
Since $H$ is translation invariant,
\begin{equation}\label{4.6}
[H,P_{\mathrm{tot}}]=0\,,
\end{equation}
there exists the direct integral decompositions corresponding to $P_{\mathrm{tot}}$ as
\begin{equation}\label{4.7}
\mathfrak{F}=\int_\mathbb{T}^{\oplus} \mathrm{d}k \mathfrak{F}(k)
\end{equation}
and
\begin{equation}\label{4.8}
H=\int_\mathbb{T}^\oplus \mathrm{d}k H(k)\,.
\end{equation}
We will need only the fiber at $k=0$. To construct $\mathfrak{F}(0)$
we consider $\hat{f}\in\mathfrak{F}$, in the momentum representation, such
that $\hat{f}_0=0$ and $\hat{f}_n$ are continuous on $\mathbb{T}^n$. Then $\hat{f}^0 \in \mathfrak{F}(0)$ is defined by
\begin{equation}\label{4.8a}
\hat{f}^0_n(k_1,\ldots,k_{n}) = \hat{f}_n(k_1,\ldots,k_{n})\delta(k_1+\ldots+k_{n})\,, \quad n = 1,2,\ldots\,.
\end{equation}
The scalar product is given by
\begin{equation}\label{4.9}
\langle \hat{f}^0_n,\hat{f}^0_n\rangle^0_n = \int_{\mathbb{T}^{n}}
\mathrm{d}k_1\ldots\mathrm{d}k_{n} \delta(k_1+\ldots+k_{n})
|\hat{f}_n(k_1,\ldots,k_{n})|^2\,,
\end{equation}
\begin{equation}\label{4.10}
\langle \hat{f}^0,\hat{f}^0\rangle^0 = \sum^\infty_{n=1}\langle \hat{f}^0_n,\hat{f}^0_n\rangle_n^0\,.
\end{equation}
$\mathfrak{F}(0)$ is the completion with respect to this norm.

The free part on $ \mathfrak{F}_n$ is multiplication by
\begin{equation}\label{4.11a}
\Omega_n(k_1,\ldots,k_n) = \sum^{n}_{\ell=1} \omega(k_\ell)\,.
\end{equation}
Hence
\begin{equation}\label{4.11}
(H_0\hat{f}^0_n)(k_1,\ldots,k_{n})= \Omega_n(k_1,\ldots,k_n)
\hat{f}_n(k_1,\ldots,k_{n})\delta(k_1+\ldots+k_{n})
\end{equation}
and $H_0\hat{f}^0 \in  \mathfrak{F}(0)$. Correspondingly, using (\ref{2.32}), the cubic operator $A$
on $\mathfrak{F}$ acts as

\begin{eqnarray}\label{4.11b}
 &&\hspace{-30pt}
(A\hat{f}_{n+1})(k_1,\ldots,k_{n})
\nonumber\\
&&\hspace{0pt}=-2i\sqrt{n+1}\Big(\sum^{n}_{\ell=1}
\int_\mathbb{T}\mathrm{d}k_{n+1}\big(\sin(2\pi k_\ell)+\sin(2\pi
k_{n+1})\big) \nonumber\\
&&\hspace{76pt}\times \hat{f}_{n+1}(k_1,\ldots,\hat{k}_\ell,\ldots,k_{n+1},
k_\ell- k_{n+1})
\end{eqnarray}
with a similar expression for $A^\ast$, see below. Clearly, if $\hat{f}_{n+1}$ contains the delta function $\delta(k_1+\ldots+k_{n+1})$, then $A\hat{f}_{n+1}$
is proportional to $\delta(k_1+\ldots+k_{n})$ and hence  $A\hat{f}^0_{n+1}  \in \mathfrak{F}_n(0)$.
Note that
$H\upharpoonright \mathfrak{F}_1(0)=0$ and
$H^\ast\upharpoonright \mathfrak{F}_1(0)=0$.

The current function at $j=0$ with its average subtracted equals  $w_0- \langle w_0\rangle =
\nu_0(u_1-u_0)+(\lambda_0/6)(u^2_0+u_0u_1+u^2_1-\alpha^{-1})$. To construct the corresponding
vector $f^w\in \mathfrak{F}$, we use
\begin{equation}\label{4.11bb}
u_j \psi_0=(2\alpha)^{-1/2} a^\ast_j \psi_0\,,\quad
\big(u^2_j-(2\alpha)^{-1}\big)\psi_0 =(2\alpha)^{-1} a^{\ast 2}_j \psi_0\,,
\end{equation}
and the fact that the ground state $\psi_0$ is mapped to the Fock vacuum. By using
(\ref{2.22b}) one obtains $f^w_n=0$ for $n=0$, $n\geq 3$, and
\begin{eqnarray}\label{4.11c}
 &&\hspace{-14pt}f^w_1(x_1)=\nu_0(2\alpha)^{-1/2}
\big(\delta(x_1-1)-\delta(x_1)\big)\,,\nonumber\\
&&\hspace{-30pt}f^w_2(x_1,x_2)=(\lambda_0/6)(2\sqrt{2}\alpha)^{-1}
\big(2\delta(x_1)\delta(x_2)+2\delta(x_1-1)\delta(x_2-1)\nonumber\\
&&\hspace{55pt}+
\delta(x_1)\delta(x_2-1)+
\delta(x_1-1)\delta(x_2)\big)\,,
\end{eqnarray}
which in Fourier space reads
\begin{eqnarray}\label{4.11d}
 &&\hspace{-14pt}\hat{f}^w_1(k_1)=\nu_0(2\alpha)^{-1/2}
(\mathrm{e}^{-i 2 \pi k_1}-1)\,,\nonumber\\
&&\hspace{-14pt}\hat{f}^w_2(k_1,k_2)=(\lambda_0/6)(2\sqrt{2}\alpha)^{-1}
\big(2+2 \mathrm{e}^{-i 2\pi (k_1+k_2)}+\mathrm{e}^{-i 2\pi k_1}+\mathrm{e}^{-i 2\pi
k_2}\big)\,.
\end{eqnarray}
$\hat{f}^{w0} \in \mathfrak{F}(0)$ is then given by
\begin{eqnarray}\label{4.11e}
&&\hspace{-14pt} \hat{f}^{w0} = \lambda_0(2\sqrt{2}\alpha)^{-1}  \hat{g}^{0}\,,\quad
 \hat{g}^{0} = (0,0,\hat{w}^0,0,0,\ldots)\,,\nonumber\\
&&\hspace{-14pt}  \hat{w}^0(k_1,k_2)=\hat{w}(k_1)\delta(k_1+k_2)\,,\quad  \hat{w}(k_1) = \tfrac{1}{3}(2+\cos(2\pi
k_1))\,,
\end{eqnarray}
enforcing  the normalization $ \hat{w}(0) = 1$.

With this input the resolvent from (\ref{4.1b}) reads
\begin{equation}\label{4.12}
\sum_{j\in\mathbb{Z}}\langle (w_0 - \langle w_0\rangle)\frac{1}{\nu_0 \zeta-L} (w_j - \langle w_j\rangle)
\rangle
=\lambda_0^2 (2\sqrt{2}\alpha)^{-2} \nu_0^{-1}\langle
\hat{g}^{0},(\zeta+H)^{-1}\hat{g}^{0} \rangle^0\,.
\end{equation}
Comparing (\ref{4.4a}) and (\ref{4.1b}) together with (\ref{4.12})
one arrives at the following prediction from the KPZ scaling theory,
\begin{equation}\label{4.13}
\langle \hat{g}^{0},(\zeta+H)^{-1}\hat{g}^{0} \rangle^0 = 3^{-2/3}\Gamma(7/3) \langle x^2\rangle_{\mathrm{KPZ}}
(\lambda^2 \zeta)^{-1/3}
\end{equation}
for small $\zeta$.
\section{Bounds and relaxation time approximation}\label{sec5}
\setcounter{equation}{0}

While at present we have no techniques to establish (\ref{4.13}),
following the methods in \cite{LQSY} one can study upper and lower bounds for the
matrix element $\langle
\hat{g}^{0},(\zeta+H )^{-1}\hat{g}^{0}\rangle^0$.
Restricting $H $ to $\bigoplus^n_{m=1} \mathfrak{F}_m(0)$
yields bounds depending on $n$, which alternate as upper and lower
bounds and which converge to a limit as $n\to \infty$.
More precisely, we define recursively
\begin{eqnarray}\label{4.16a}
&&\hspace{-30pt}T_2 = (\zeta+H_0)^{-1}\,, \nonumber\\
&& \hspace{-30pt}T_n = (\zeta+H_0+\lambda^2 A T_{n-1} A^*)^{-1}\,,
\end{eqnarray}
and set
\begin{equation}
 b_n(\zeta) = \langle \hat{g}^{0}, T_n  \hat{g}^{0} \rangle^0\,.
\end{equation}
E.g.,
\begin{eqnarray}\label{4.16b}
 && \hspace{-40pt} b_3(\zeta) =  \langle\hat{g}^{0}\,,
 \big\{ \zeta+H_0  + \lambda^2 A  (\zeta+H_0 )^{-1}A ^*\big\}^{-1}
 \hat{g}^{0}\rangle^0\,, \nonumber\\
 && \hspace{-40pt} b_4(\zeta) =  \langle\hat{g}^{0}\,,
 \big\{ \zeta+H_0  + \lambda^2 A
 \big(\zeta+H_0  +\lambda^2 A (\zeta+H_0 )^{-1}A ^*\big)^{-1}A ^*\big\}^{-1}
 \hat{g}^{0}\rangle^0\,.
\end{eqnarray}
Then
\begin{equation}
 b_3(\zeta) \leq b_5(\zeta) \leq \ldots \leq b_4(\zeta) \leq b_2(\zeta)\,,
 \end{equation}
 and
 \begin{equation}
 \langle \hat{g}^{0}, (\zeta+H  )^{-1} \hat{g}^{0}\rangle^0
 = \lim_{n\to\infty} b_n(\zeta)\,.
\end{equation}

The vector $ \hat{g}^{0}$ has $\hat{g}_2^{0} = \hat{w}^0 $ as the  only nonvanishing entry.
$H_0$ preserves the particle number, while $A$ decreases it by one
and $A^*$ increases it by one. Hence it will be convenient to introduce a notation
displaying these special features.
Let $P_n $ be the projection onto $\mathfrak{F}_n$. We set
\begin{equation}\label{4.18}
P_n  H_0  P_n =H_{0,n}
\end{equation}
and
\begin{equation}\label{4.19}
P_n  A  P_{n+1} =A_{n,n+1} \,,\quad
 P_{n+1}  A ^\ast P_n =A_{n+1,n} \,.
\end{equation}
 For a given $n$, we define recursively
\begin{equation}\label{4.18a}
U^{(n)}_{n-1} = \lambda^2 A_{n-1,n}(\zeta+H_{0,n})^{-1}A_{n,n-1}\,,
\end{equation}
 \begin{equation}\label{4.18b}
 U^{(n)}_m     = \lambda^2 A_{m,m+1}(\zeta+H_{0,m+1}+U^{(n)}_{m+1})^{-1}A_{m+1,m}\,, \quad 2\leq m\leq n-2\,.
\end{equation}
$U^{(n)}_m$ acts on $\mathfrak{F}_m$ and leaves $\mathfrak{F}_m(0)$ invariant. Our bound $b_n(\zeta)$ can then be written as
\begin{equation}
 b_n(\zeta) = \langle \hat{w}^0, (\zeta+H_{0,2}+U^{(n)}_2 )^{-1} \hat{w}^0\rangle_2^0\,.
\end{equation}

The next task is to work out more concretely the operator
$A_{n,n+1}(\zeta+H_{0,n+1})^{-1}$ $A_{n+1,n}$ appearing in (\ref{4.18a}).
We use duality to compute $A^\ast$,
\begin{eqnarray}\label{4.19a}
 && \hspace{-10pt}\langle f_n, A f_{n+1} \rangle_n =
\int_{\mathbb{T}^n} \mathrm{d}\underline{k} f_n(k_1,\ldots,k_n)^\ast A f_{n+1}(k_1,\ldots,k_n)\nonumber\\
 && \hspace{10pt} = -2i \sqrt{n+1}\sum^n_{\ell=1}
\int_{\mathbb{T}^{n+1}} \mathrm{d}\underline{k} \mathrm{d}k_{n+1}\big(\sin (2\pi
k_\ell)+\sin(2\pi k_{n+1})\big) f_n(k_1,\ldots,k_n)^\ast\nonumber\\
 && \hspace{100pt}\times  f_{n+1} (k_1,\ldots,k_\ell-k_{n+1},\ldots,
k_{n+1})\nonumber\\
 && \hspace{15pt}=\int_{\mathbb{T}^{n+1}} \mathrm{d}\underline{k}\mathrm{d}k_{n+1}f_{n+1}(k_1,\ldots,k_{n+1})
\Big(2i \sqrt{n+1}\sum^n_{\ell=1}\big(\sin(2\pi(k_\ell+k_{n+1}))\nonumber\\
 && \hspace{100pt} +\sin(2\pi k_{n+1})\big)f_n (k_1,\ldots,k_\ell+k_{n+1},\ldots,
k_{n})\Big)^\ast\,.
\end{eqnarray}
Upon symmetrization, one arrives at
\begin{eqnarray}\label{4.19b}
 &&\hspace{-21pt}A^\ast f_n(k_1,\ldots,k_{n+1})= 2i \frac{1}{\sqrt{n+1}}
\sum_{1\leq j<\ell\leq n+1}\big(2 \sin (2\pi(k_j+k_\ell))+\sin(2\pi
k_j)\nonumber\\
 &&\hspace{95pt}+\sin(2\pi k_\ell)\big)f_n(k_1,\ldots,\hat{k}_j,\hat{k}_\ell,\ldots,k_j+k_\ell) \,.
\end{eqnarray}
To compute $U^{(n+1)}_n$ we
denote the shift $k_i\rightsquigarrow k_i-k_{n+1}$ by
$\tau_i$. Then
\begin{eqnarray}\label{4.19d}
 && \hspace{-30pt}U^{(n+1)}_n f_n (k_1,\ldots,k_n) = \lambda^2 (A_{n,n+1}(\zeta+H_{0,n+1})^{-1} A_{n+1,n}
f_n)(k_1,\ldots,k_n)\nonumber\\
 && \hspace{-10pt}= 4 \lambda^2\sum^n_{i=1} \sum_{1\leq j<\ell\leq n+1}
\int_\mathbb{T} \mathrm{d}k_{n+1}\big(\sin(2\pi(k_i - k_{n+1}))+\sin(2\pi k_{n+1})\big)\nonumber\\
&& \hspace{20pt}\times(\zeta+\Omega_{n+1}(k_1,\ldots,k_\ell - k_{n+1},\ldots,k_{n+1}))^{-1}
\tau_i\big(\sin(2\pi k_j)+\sin(2\pi
k_\ell)\nonumber\\
 && \hspace{50pt} +2\sin(2\pi(k_j+k_\ell))\big)\tau_i f_n (k_1,\ldots,\hat{k}_j,\hat{k}_\ell,\ldots,
k_j+k_\ell)\,.
\end{eqnarray}
Clearly, $U^{(n+1)}_n=V^{(n+1)}_n+R^{(n+1)}_n$, where $V^{(n+1)}_n$ is a multiplication operator and
$R^{(n+1)}_n$ an integral operator. In the sum in (\ref{4.19d}), $V^{(n+1)}_n$
corresponds to the terms $i=j$ and $\ell=n+1$. Hence
\begin{eqnarray}\label{4.19e}
 &&\hspace{-25pt}V^{(n+1)}_n(k_1,\ldots,k_n)= 2 \lambda^2\sum^n_{\ell=1}
\int_\mathbb{T}\mathrm{d}k_{n+1} \big( \sin (2\pi(k_\ell-k_{n+1}))+\sin(2\pi
k_{n+1})\nonumber\\
 &&\hspace{35pt} +2\sin(2\pi k_\ell)\big)^2 (\zeta+\Omega_{n+1}(k_1,\ldots,k_\ell - k_{n+1},\ldots,
 k_{n+1}))^{-1}\,.
\end{eqnarray}

Considering the special cases $n=2,3$ one arrives at the following bounds.
\begin{theorem}\label{t1}
In the limit $\zeta\to 0$, the following bounds
are valid,
\begin{equation}\label{4.17}
\lambda^{-1} 2^{-5/4}  3^{-3/2} \zeta^{-1/4}\leq \langle
\hat{g}^{0},(\zeta+H )^{-1} \hat{g}^{0}\rangle^0 \leq   2^{-3/2}
 \zeta^{-1/2}\,.
\end{equation}
\end{theorem}
While the lower bound is not sharp, it establishes that $S(j,t)$
must spread superdiffusively.\medskip\\
\begin{proof}
$n=2$: It holds
\begin{eqnarray}\label{4.20}
 &&\hspace{-40pt}\langle\hat{g}^{0},\big(\zeta+H \big)^{-1}\hat{g}^{0}\rangle^0\leq
\langle\hat{w}^0,\big(\zeta+H_{0,2} \big)^{-1}\hat{w}^0\rangle^0_2\nonumber\\
&&\hspace{-10pt} =\int_\mathbb{T}\mathrm{d}k_1
\big(\tfrac{1}{3}(2+\cos(2\pi k_1))\big)^2
\big(\zeta+2\omega(k_1)\big)^{-1}\nonumber\\
&&\hspace{-10pt}
\cong  2^{-3/2} \zeta^{-1/2}
\end{eqnarray}
for $\zeta\to 0$.\medskip\\
$n=3$: It holds
\begin{eqnarray}\label{4.21}
 &&\hspace{-35pt}\langle\hat{g}^{0},\big(\zeta+H \big)^{-1}\hat{g}^{0}\rangle^0
 \nonumber\\
&&\hspace{-25pt}\geq
\langle\hat{w}^0,\big[\zeta+H_{0,2} +\lambda^2 A_{2,3}  \big(\zeta+H_{0,3} \big)^{-1} A_{3,2}  \big]^{-1}
\hat{w}^0\rangle_2^0\,.
\end{eqnarray}
For the quadratic form of $\lambda^2 A_{2,3}  \big(\zeta+H_{0,3} \big)^{-1} A_{3,2}$ one obtains, with
$\hat{f}_2\in \mathfrak{F}_2$,
\begin{eqnarray}\label{4.22}
 &&\hspace{-25pt}\lambda^2\langle \hat{f}_2, A_{2,3}(\zeta+H_{0,3})^{-1} A_{3,2}
\hat{f}_2\rangle_2= (4\lambda^2/3)\int_{\mathbb{T}^3} \mathrm{d}k_1
\mathrm{d}k_2\mathrm{d}k_3
\big(\zeta+\Omega_3(k_1,k_2,k_3)\big)^{-1}\nonumber\\
&&\hspace{22pt}\Big(\big(2 \sin(2\pi(k_1+k_2))+\sin(2\pi k_1)+
\sin(2\pi
k_2)\big)\hat{f}_2(k_1+k_2,k_3)\nonumber\\
&&\hspace{34pt}+\big(2 \sin(2\pi(k_1+k_3))+\sin(2\pi k_1)+ \sin(2\pi
k_3)\big)\hat{f}_2(k_1+k_3,k_2)\nonumber\\
&&\hspace{34pt}+\big(2 \sin(2\pi(k_2+k_3))+\sin(2\pi k_2)+ \sin(2\pi
k_3)\big)\hat{f}_2(k_2+k_3,k_1)\Big)^2\nonumber\\
&&\hspace{34pt}\leq 3\langle \hat{f}_2, V^{(3)}_2 \hat{f}_2\rangle_2\,,
\end{eqnarray}
where Schwarz inequality is used in the last step. By taking limits, the inequality holds also for the fiber $k=0$. Hence
\begin{equation}\label{4.24}
\langle\hat{g}^{0},(\zeta+H )^{-1}\hat{g}^{0}\rangle^0\geq \langle \hat{w}^0,
(\zeta+H_{0,2} + 3 V^{(3)}_2 )^{-1}
\hat{w}^0\rangle^0_2\,.
\end{equation}
For small $\zeta$ the dominant part of the integral comes from $k_1$ close to 0 and one
obtains
\begin{eqnarray}\label{4.25}
 &&\hspace{-35pt}\langle\hat{w}^0,(\zeta+H_{0,2} +3
V^{3}_2 )^{-1}\hat{w}^0\rangle^0_2\nonumber\\
&&\hspace{-20pt}\cong \int_\mathbb{T}\mathrm{d}k_1 (\zeta+(2\pi
k_1)^2 \lambda^2 2^{1/2} 3^3 \zeta^{-1/2})^{-1} = 2^{-1} (\lambda^2
2^{1/2} 3^3 )^{-1/2} \zeta^{-1/4}\,.
\end{eqnarray}
\end{proof}

It is of interest to understand whether Theorem \ref{t1} could be improved to
yield the KPZ exponent 1/3 as
$n\to \infty$. Unfortunately, already for $n=4$, one would need a lower
bound on $A_{3,4}(\zeta+H_{0,4})A_{4,3}$ by a multiplication operator,
compare with (\ref{4.22}). Such a bound does not seem to be
available. To make, nevertheless, some progress we observe the splitting
\begin{equation}\label{pot_app}
U^{(n)}_{n-1} = V^{(n)}_{n-1} + R^{(n)}_{n-1}
\end{equation}
in (\ref{4.19d}), where $V^{(n)}_{n-1}$ is a multiplication operator and the remainder $R^{(n)}_{n-1}$ is
an integral operator involving either a single particle or a pair
of particles. Therefore one expects that the limit $\zeta \to 0$
will be dominated by the potential $V^{(n)}_{n-1}$.
In solid state physics this type of approximation is called the
relaxation time approximation. Of course it is uncontrolled,
at least at present.

 With this approximation in the expression for $U^{(n)}_{n-2}$ one substitutes $V^{(n)}_{n-1}$
 for $U^{(n)}_{n-1}$. Again for $U^{(n)}_{n-2}$ one neglects the integral operator
 $R^{(n)}_{n-2}$. Iterating this procedure yields the recursion relation for the potentials $V^{(n)}_m$,
 \begin{eqnarray}
&&\hspace{-20pt} V^{(n)}_{n} = 0\,,\nonumber\\[1ex]
 &&\hspace{-20pt}V^{(n)}_m(k_1,\ldots,k_m) \nonumber\\
 &&\hspace{-10pt}=
 2 \lambda^2  \sum_{\ell=1}^m \int_{\mathbb{T}}  \mathrm{d} k_{m+1}
 \big(2\sin (2\pi  k_\ell)  + \sin (2\pi (k_\ell-k_{m+1})) + \sin (2\pi k_{m+1})\big)^2 \nonumber\\
 &&\hspace{-10pt}
     \times \big(\zeta + \Omega_{m}(k_1,\ldots,k_\ell - k_{m+1},\dots,k_{m+1})
             + V^{(n)}_{m+1}(k_1,\ldots,k_\ell -k_{m+1},\ldots,k_{m+1})\big)^{-1}\,,\nonumber\\ [1ex]
 &&\hspace{260pt} \quad 2\leq m\leq n-1\,. \label{Vm}
 \end{eqnarray}
 Denoting $b_n(\zeta)$ within this approximation by $d_n(\zeta)$, one arrives at
 \begin{equation}
 \hspace{-20pt}d_n(\zeta)
 =
 \int_{\mathbb{T}} \mathrm{d}k \frac{|\hat{w}(k)|^2}
 {\zeta+2\omega(k)+V_2(k,-k)}\,.
 \label{bnint}
\end{equation}
Here $\hat{w}(k)$ is defined in (\ref{4.11e}), $\hat{w}(0) = 0$.

For small $\zeta$ the dominant contribution to the integral (\ref{Vm})
comes from $k_{m+1}$ close to 0. Let us start with $m=n-1$. Then
\begin{eqnarray}
 &&\hspace{-10pt}V^{(n)}_{n-1}(k_1,\ldots,k_{n-1})\cong
 (2 \lambda^2)9   \Omega_{n-1}(k_1,\ldots,k_{n-1})\int_\mathbb{T}\mathrm{d}k_n
(\zeta+2\omega(k_n))^{-1}\nonumber\\ [1ex]
 &&\hspace{86pt} \cong 9 \lambda^2 ( 2 \zeta)^{-1/2}
\Omega_{n-1} (k_1,\ldots, k_{n-1})\,. \label{5.26}
 \end{eqnarray}
The next iteration reads
\begin{eqnarray}
 &&\hspace{-10pt}V^{(n)}_{n-2}(k_1,\ldots,k_{n-2})\nonumber\\
 &&\hspace{15pt} \cong
( 2 \lambda^2)  9
\Omega_{n-2}(k_1,\ldots,k_{n-2})\int_\mathbb{T}\mathrm{d}k_{n-1}
(\zeta+(9\lambda^2)2\omega(k_{n-1})(2\zeta)^{-1/2})^{-1}\nonumber\\
 &&\hspace{15pt} \cong (9 \lambda^2)^{1/2} ( 2 \zeta)^{-1/4}
\Omega_{n-2} (k_1,\ldots, k_{n-2})\,, \label{5.27}
 \end{eqnarray}
and, in general,
\begin{eqnarray}
 &&\hspace{-10pt}V^{(n)}_{m}(k_1,\ldots,k_{m}) \cong
(9\lambda^2)^{1-\alpha_{n-m}} ( 2 \zeta)^{-\alpha_{n-m+1}} \Omega_m
(k_1,\ldots, k_{m})\,,\nonumber\\[1ex]
 &&\hspace{210pt} 2\leq m\leq n-1 \,, \label{5.28}
 \end{eqnarray}
where the exponents $\alpha_j$ are defined recursively through
\begin{equation}\label{5.29}
 \alpha_{j+1}=\tfrac{1}{2}(1-\alpha_j)\,,\quad \alpha_1=0\,.
\end{equation}
Substituting (\ref{5.28}) in (\ref{bnint}) and using $\hat{w}(0)=1$
yields for small $\zeta$
\begin{equation}\label{5.30}
d_n(\zeta)= 2^{-1}(9\lambda^2)^{-\alpha_{n-1}}(2\zeta)^{-\alpha_n}\,.
\end{equation}

The solution to (\ref{5.29}) reads
\begin{equation}\label{5.31}
 \alpha_n=\frac{1}{3}\big(1-(-2)^{-(n-1)}\big)\,,
\end{equation}
which for $n\to\infty$ converges to $\alpha_\infty=1/3$. Therefore
$\lim_{n\to\infty} d_n(\zeta)=d_\infty(\zeta)$ and
\begin{equation}\label{5.32}
 d_\infty(\zeta)=2^{-4/3} 3^{-2/3}(\lambda^2 \zeta)^{-1/3}\,,
\end{equation}
which should be compared with (\ref{4.13}). Remarkably enough, the
relaxation time approximation yields the KPZ exponent $1/3$. The
prefactor of $(\lambda^2 \zeta)^{-1/3}$ equals 0.292 in (\ref{4.13})
while it is 0.191 in (\ref{5.32}). Thus the relaxation time
approximation gives a prefactor which is approximately $2/3$ off the
true value. We take this as an indication that one needs more
powerful methods to obtain the universal scaling form for the
two-point function.

\section{Continuum limit}\label{sec6}
\setcounter{equation}{0}

As remarked in the Introduction, the physically meaningful continuum
limit of (\ref{1.3}) must be such as to preserve the scaling
(\ref{4.3}) of the two-point function. Here we want to amplify this
point and start on the unit lattice. To avoid the various constants,
we make the specific choice $ \nu_0=1/2$, $D_0=1$. Let us first
consider the linear case, $\lambda_0=0$. Then on the microscopic
scale
\begin{equation}\label{6.1}
\frac{d}{dt}
u_j(t)=\tfrac{1}{2}\big(u_{j+1}(t)-2u_j(t)+u_{j-1}(t)\big)
+\xi_j(t)-\xi_{j-1}(t)\,.
\end{equation}
For the continuum limit we average over a smooth test function $g$,
$g:\mathbb{R}\to\mathbb{R}$, varying on the scale $\delta^{-1}$,
$\delta\ll 1$, and take long times $\delta^{-2}t$ with
$t=\mathcal{O}(1)$. Then
\begin{equation}\label{6.2}
\lim_{\delta\to 0} \delta \sum_{j\in\mathbb{Z}} g(\delta j)
\delta^{-1/2} u_j (\delta^{-2}t)= \int \mathrm{d}x g(x) \phi (x,t)\,,
\end{equation}
where $\phi$ is a mean zero Gaussian field with covariance
\begin{equation}\label{6.3}
\langle \phi (x,t) \phi(x',t')\rangle=(2\pi|t-t'|)^{-1/2}
\exp\big[-(x-x')^2/2|t-t'|\big]\,.
\end{equation}

Equivalently one can view $u_j$ on the lattice with spacing $\delta$
by defining
\begin{equation}\label{6.3a}
v^\delta(x,t)=\delta^{-1/2} u_{\lfloor x/\delta\rfloor}
(\delta^{-2} t)
\end{equation}
 with $\lfloor\cdot\rfloor$ denoting integer part and
$x\in\mathbb{R}$. Then, integrating against smooth test functions,
the continuum limit reads
\begin{equation}\label{6.4}
\lim_{\delta\to 0} v^\delta(x,t)=\phi (x,t)\,.
\end{equation}
By (\ref{6.1}) $v^\delta$ satisfies
\begin{eqnarray}\label{6.5}
 &&\hspace{-47pt}\frac{d}{dt}v^\delta(x,t)=\delta^{-2} \tfrac{1}{2}
\big(v^\delta(x+\delta,t)-2v^\delta(x,t)+v^\delta(x-\delta,t)\big)\nonumber\\
&&\hspace{18pt} +\delta^{-1/2} \delta^{-1}\big(\xi_{\lfloor
x/\delta\rfloor}(t)-
\xi_{\lfloor(x-\delta)/\delta\rfloor}(t)\big)\,.
\end{eqnarray}
Comparing with (\ref{1.3}), we observe that the bare coefficients
are not rescaled. Taking the limit $\delta\to 0$ in (\ref{6.5})
yields
\begin{equation}\label{6.5a}
\partial_t\phi(x,t)=\tfrac{1}{2} \partial^2_x \phi(x,t)
+\partial_x\xi(x,t)
\end{equation}
with white noise initial data, in agreement with (\ref{6.3}),
(\ref{6.4}).

Next we include the nonlinearity with $\lambda_0=1$. Then
\begin{equation}\label{6.6}
\frac{d}{dt}u_j(t)=\tilde{w}_j(t)-\tilde{w}_{j-1}(t)
+\tfrac{1}{2}\big(u_{j+1}(t) -2u_j(t)
+u_{j-1}(t)\big)+\xi_j(t)-\xi_{j-1}(t)
\end{equation}
and the stationary two-point function should scale as
\begin{equation}\label{6.7}
\lim_{\delta\to 0} \delta S([\delta^{-1}x],\delta^{-3/2}t)= t^{-2/3}
f_{\mathrm{KPZ}}(t^{-2/3}x)\,.
\end{equation}
In particular, the correct time scale for the continuum limit is
$\delta^{-3/2}t$, $t=\mathcal{O}(1)$. In analogy to (\ref{6.4}) one
introduces the macroscopic field $v^\delta(x,t)$ by
\begin{equation}\label{6.8}
v^\delta(x,t)=\delta^{-1/2} u_{\lfloor x/\delta\rfloor}
(\delta^{-3/2} t)\,.
\end{equation}
According (\ref{6.6}), $v^\delta(x,t)$ is governed by the evolution
\begin{eqnarray}\label{6.9}
 &&\hspace{-47pt}\frac{d}{dt}v^\delta(x,t)=
\delta^{-1}\big(\tilde{w}^\delta(x,t)-\tilde{w}^\delta(x-\delta,t)\big)\nonumber\\
&&\hspace{18pt} +\delta^{1/2}
\delta^{-2}\tfrac{1}{2}\big(v^\delta(x+\delta,t) -2v^\delta(x,t)
+v^\delta(x-\delta,t)\big)\nonumber\\
&&\hspace{18pt} + \delta^{1/4} \delta^{-1/2} \delta^{-1}
\big(\xi_{\lfloor x/\delta\rfloor}(t) -\xi_{\lfloor
(x-\delta)/\delta\rfloor}(t)\big)\,.
\end{eqnarray}
Comparing with (\ref{1.3}) we conclude that the nonlinearity is left
invariant, thus $\lambda_\mathrm{b}$ remains fixed, while the linear
part is scaled down by $\sqrt{\delta}$. Of course, it is natural to
conjecture that, as in the linear case, $v^\delta(x,t)$ has a limit
as $\delta\to 0$. To identify the limit one could also use other,
better understood models, like the TASEP. Some properties are known
\cite{FS06,BFP09}, but the full limit of $v^\delta(x,t)$ still has
to be identified.

There is one particular case for which analytical results are
available \cite{FrMa,ChDu}. In fact, it is the problem Burgers
wanted to solve \cite{Bur}. In (\ref{1.2}) one sets $D_\mathrm{b}=0$
and studies the decay of the solution for Gaussian white noise
initial data, $ \langle u(x,0) u(x',0)\rangle$ $= (1/8)\delta(x-x')$.
We set $\lambda_\mathrm{b}=1$. The solution to (\ref{1.2}) is well
defined in the limit $\nu_\mathrm{b}=0$. With this meaning, we set
$\nu_\mathrm{b}=0$. Then the solution $u(x,t)$ is statistically
self-similar, in the sense that
\begin{equation}\label{6.9a}
u(x,t)=t^{-1/3} u(t^{-2/3} x,1)
\end{equation}
in distribution. $x\to u(x,1)$ is a stationary Markov process with
the generator
\begin{equation}\label{6.10}
L_\mathrm{T} f(u)=\frac{d}{du} f(u)+\int^u_{-\infty} \mathrm{d}u'
R(u,u')\big(f(u')-f(u)\big)
\end{equation}
for functions $f:\mathbb{R}\to\mathbb{R}$. The second term
corresponds to a Markov jump process. Thus $u(x,1)=a+x$ locally,
interrupted by downward jumps from $u$ to $u'$ with rate $R(u,u')$.
The rate function is computed explicitly and given by
\begin{equation}\label{6.11}
R(u,u')=(u-u')\frac{J(u')}{J(u)} I(u-u')\,,
\end{equation}where $I$ and $J$ are given by their Fourier and
Laplace transforms in terms of the Airy function Ai,
\begin{equation}\label{6.12}
J(u)= \frac{1}{2\pi i}\int^{i\infty}_{-i\infty} \mathrm{d}z
\frac{\exp (uz)}{2^{1/3} \mathrm{Ai}(2^{-1/3}z)}\,,
\end{equation}
\begin{equation}\label{6.13}
2I(u)= (2\pi
u^3)^{-1/2}+\frac{1}{2\pi i}\int^{i\infty}_{-i\infty}\mathrm{d}z
\exp(uz) \Big(\frac{2^{2/3}
\mathrm{Ai'}(2^{-1/3}z)}{\mathrm{Ai}(2^{-1/3}z)}+ (2z)^{1/2}\Big)\,.
\end{equation}

Plots of the stationary distribution for $L_T$, and other
quantities, can be found in \cite{FrMa}.

\end{document}